\documentclass[conference]{IEEEtran}
\usepackage{ragged2e}
\usepackage{floatrow}
\usepackage{amsthm}
\usepackage[utf8]{inputenc}
\usepackage{soul}            % for \st (strikeout command)
\usepackage[english]{babel}
\theoremstyle{remark}
\newtheorem{theorem}{Theorem}

\newtheorem{Proposition}{Proposition}

\usepackage{multicol}
\usepackage{epsfig}
\usepackage{epstopdf}
\usepackage{float}
\usepackage{perpage}
\MakeSorted{figure}
\MakeSorted{table}
\usepackage{array}
\usepackage{stfloats}
\usepackage{graphicx}
\usepackage{amsfonts}
\usepackage{amssymb}

\usepackage{enumitem}
\usepackage{verbatim}
\usepackage[english]{babel}
\usepackage[utf8]{inputenc}
\usepackage[linesnumbered,ruled,vlined]{algorithm2e}

\usepackage{cite}
\DeclareMathAlphabet\mathbfcal{OMS}{cmsy}{b}{n}
\ifCLASSINFOpdf
\else
\fi
\usepackage[cmex10]{amsmath}
\usepackage{array}
\usepackage{fixltx2e}
\newcounter{mytempeqncnt}
\usepackage[linesnumbered,ruled,vlined]{algorithm2e}
\usepackage[font=footnotesize,labelfont=bf, figurename=Fig.]{caption} %scriptsize\\
\usepackage{subcaption}
\usepackage{fix-cm}
\usepackage{etoolbox} % for \iftoggle
\newtoggle{SINGLE_COL}
\togglefalse{SINGLE_COL}   % for two column format

\newtoggle{BIG_EQUATION}
\toggletrue{BIG_EQUATION}  % for two column format


\usepackage{mathtools}

\newfloatcommand{capbtabbox}{table}[][\FBwidth]

\newcommand{\Amat}{\mathbf{A}}
\newcommand{\Bmat}{\mathbf{B}}
\newcommand{\Cmat}{\mathbf{C}}
\newcommand{\Dmat}{\mathbf{D}}

\newcommand{\Imat}{\mathbf{I}}

\newcommand{\Rmat}{\mathbf{R}}

\newcommand{\Tmat}{\mathbf{T}}

\newcommand{\Xmat}{\mathbf{X}}

\newcommand{\cv}{\mathbf{c}}

\newcommand{\gv}{\mathbf{g}}
\newcommand{\hv}{\mathbf{h}}

\newcommand{\nv}{\mathbf{n}}
\newcommand{\pv}{\mathbf{p}}
\newcommand{\qv}{\mathbf{q}}
\newcommand{\rv}{\mathbf{r}}

\newcommand{\tv}{\mathbf{t}}
\newcommand{\uv}{\mathbf{u}}
\newcommand{\vv}{\mathbf{v}}

\newcommand{\xv}{\mathbf{x}}
\newcommand{\yv}{\mathbf{y}}
\newcommand{\zv}{\mathbf{z}}

\newcommand\delequal{ \overset{\Delta}{=}}
\newcommand{\Thetamat}{\boldsymbol{\Theta}}

\newcommand{\Psimat}{\boldsymbol{\Psi}}

\newcommand{\etamat}{\boldsymbol{\eta}}
\newcommand{\phimat}{\boldsymbol{\phi}}

\newcommand{\E}{\mathbb{E}}

\SetKwInput{KwInput}{Input}
\SetKwInput{KwOutput}{Output}
\begin{document}
%	\title{Receiver Design for Hardware-Impaired Multi-cell Multi-IRS mMIMO Systems  and Channel Aging}
	\title{Multi-cell mMIMO IRS Systems With  Impairments and Aging: Phase Optimization and Receiver Design}
%	\title{IRS-Aided Multi-cell mMIMO Systems: Hardware Impairments and Aging Aware Receiver Design}
%\title{Receiver Design  for IRS-Aided Multi-cell  Massive MIMO Systems With Channel Aging and Non-Ideal Hardware}
\author{\\[-20pt]\IEEEauthorblockN{Rakesh Munagala, Dheeraj Naidu Amudala, Rohit Budhiraja}
\IEEEauthorblockA{\textit{Department of Electrical Engineering, IIT Kanpur India}\\ 
\{mrakesh, dheeraja, rohitbr\}@iitk.ac.in}\\[-40pt]}
\maketitle
\begin{abstract}
	 We consider the uplink of a hardware-impaired intelligent reflective surfaces (IRS) aided multi-cell massive multiple-input multiple-output (mMIMO) system with mobile user equipments, whose channel age with time. For this system, we design a novel distortion-and-aging-aware MMSE (DAA-MMSE) receiver that not only provides a higher spectral efficiency (SE) than  conventional maximal ratio and distortion-unaware MMSE (DU-MMSE) receivers, but also reduces the pilot overhead. We develop a novel low-complexity IRS phase optimization framework based on minorization-maximization (MM) technique, which requires only channel statistics to calculate the optimal phase. We also show that the SE gain of the  DAA-MMSE receiver over DU-MMSE receiver increases with hardware impairments, and channel aging. Along with DAA-MMSE receiver, the IRS is also shown to reduce the pilot overhead in a mMIMO system with channel aging.
% We consider the uplink of a hardware-impaired intelligent reflective surfaces (IRS) aided multi-cell massive multiple-input multiple-output (mMIMO) system with mobile user equipments, whose channel age with time. For this system, we analyze the distortion due to hardware impairments, and the effect of channel aging, and propose a novel distortion-and-aging-aware MMSE (DAA-MMSE) receiver that not only provides a higher spectral efficiency (SE) than  conventional maximal ratio and distortion-unaware MMSE (DU-MMSE) receivers, but also reduces the pilot overhead. We develop a novel low-complexity SE optimization framework based on minorization-maximization (MM) technique, which requires only channel statistics to find the optimal phase. We also show that the SE gain of the  DAA-MMSE receiver over DU-MMSE receiver increases with hardware impairments, and channel aging. Along with DAA-MMSE receiver, the  IRS is also shown to reduce the pilot overhead in a mMIMO system with channel aging.
\end{abstract}
\vspace{-8pt}
\section{Introduction}\vspace{-3pt}
Massive multiple-input multiple-output (mMIMO) is a key technology for the current fifth generation (5G) cellular systems~\cite{bjornson2018hardware}. With attenuated signal due to high-rise buildings and poor scattering conditions, even a mMIMO base station (BS) cannot guarantee adequate coverage and quality-of-service (QoS)~\cite{xie2020max,zhi2022ris}. Intelligent reflective surface (IRS)~\cite{xie2020max,zhi2022ris} technology is being investigated to improve these aspects by directly modifying the propagation channel. An IRS consists of a large number of reflecting elements, which by shifting the phase of the incident signal, can favorably modify the  UE-to-BS  channels~\cite{xie2020max,zhi2022ris}. Xie~\textit{et al.} in~\cite{xie2020max} maximized the instantaneous signal-to-interference-plus-noise-ratio (SINR) of downlink multi-cell system by optimizing transmit beamforming vector, and IRS phase matrix using second-order-cone programming and successive convex approximation, respectively. The authors in~\cite{zhi2022ris} investigated the performance of IRS-aided single-cell mMIMO system by using the zero-forcing detector.

The cellular systems beyond 5G are being designed to support a UE velocity of up to $500$~km/h~\cite{papazafeiropoulos2023impact}. The UE mobility causes the channel to continuously evolve in time. The channel estimated by the BS, consequently, becomes out-dated, a phenomenon known as channel aging~\cite{papazafeiropoulos2023impact,zhang2023channel}. The authors in~\cite{papazafeiropoulos2023impact} analyzed the SE of IRS-aided mMIMO system with channel aging and  correlated Rayleigh fading channels. Zhang \textit{et al.} in~\cite{zhang2023channel} proposed an aging-aware combiner that depends on the instantaneous channels for an IRS-assisted mMIMO system with correlated Rayleigh fading channels. The  mMIMO works in~\cite{papazafeiropoulos2023impact,zhang2023channel}, however, 
considered a \textit{single-cell} system. Practical cellular systems deployments are multi-cell, where a UE experiences multi-user interference (MUI) from UEs in its own and neighboring cells~\cite{cai2022irs,buzzi2021ris}. The multi-cell IRS works in \cite{cai2022irs,buzzi2021ris}, used either perfect channel knowledge~\cite{cai2022irs} or separately estimated UE-to-IRS and IRS-to-BS channels~\cite{buzzi2021ris}. Individual channel estimation has  a high estimation overhead, which makes it unsuitable for aging channels~\cite{papazafeiropoulos2023impact}. \textit{The authors in~\cite{cai2022irs,buzzi2021ris} also considered only Rayleigh channels, and  that too without channel aging.}

A Rayleigh-fading channel models only the non-line-of-sight (NLoS) paths, while a Rician-fading channel models both
line-of-sight (LoS) and NLoS paths, and accurately characterizes the propagation environment~\cite{pan2020multicell,hua2020intelligent}. The 5G standardization body also evaluates potential mMIMO technologies using  Rician-fading channels before incorporating them in the standard~\cite{3gpp25.996}. This motivates us  to study, similar to~\cite{3gpp25.996,pan2020multicell,hua2020intelligent},  multi-cell IRS-aided mMIMO systems with Rician channels. Pan~\textit{et al.} in~\cite{pan2020multicell} analyzed the weighted sum SE of such system,  while Hua \textit{et al.} in~\cite{hua2020intelligent} optimized the transmit precoder at the BS and phase shifts at the IRS to guarantee fairness among the UEs. 
%These Rician-faded multi-cell IRS works in \cite{pan2020multicell, hua2020intelligent} used instantaneous SE for IRS phase and precoder/combiner designs. The  IRS phase, which is designed using instantaneous CSI, causes high CSI estimation overhead.   Luo \textit{et al}. in~\cite{luo2021reconfigurable} derived an approximate upper bound on the downlink sum SE. Jiang \textit{et al.} in~\cite{jiang2023joint} used the derived approximate closed-form SE expression for designing the beamformer that maximizes the downlink sum SE.  
\textit{The aforementioned multi-cell Rician-fading works in~\cite{pan2020multicell,hua2020intelligent} did not consider channel aging.} 
%Also, these works either maximized the instantaneous sum SE~\cite{pan2020multicell,hua2020intelligent,cai2022irs,buzzi2021ris} or approximated it~\cite{luo2021reconfigurable,jiang2023joint} to design the combiner/precoder. 
Further, the IRS-aided mMIMO works with channel aging~\cite{papazafeiropoulos2023impact,zhang2023channel} or without channel aging~\cite{pan2020multicell,hua2020intelligent} assumed high-quality radio frequency (RF) transceivers and high-resolution analog-to-digital converters (ADCs)/digital-to-analog converters (DACs) at the BS and UEs. The RF transceivers used to design practical 5G mMIMO cellular systems  have inherent hardware distortions, which degrade the system SE~\cite{papazafeiropoulos2021intelligent,peng2022performance}.
% Guo \textit{et al.} in~\cite{guo2021performance} investigated IRS-aided satellite-unmanned aerial vehicle terrestrial networks, by considering RF impairments. They showed that IRS improves the link quality, and consequently the outage probability. 
{The authors in~\cite{papazafeiropoulos2021intelligent,peng2022performance} analyzed a single-cell IRS-assisted hardware-impaired mMIMO system with Rayleigh channels, and that too without channel aging. Also, they did not design receivers to mitigate the degradation caused due to hardware impairments.}

The IRS-mMIMO literature, with/without channel aging, either considered hardware-impaired systems with maximum-ratio combining (MRC) at the BS~\cite{papazafeiropoulos2021intelligent}, or optimized the precoders by assuming ideal hardware~\cite{cai2022irs,buzzi2021ris}.
% {The authors of~\cite{lin2022slnr} and~\cite{lin2021secrecy} developed secure and energy-efficient beamforming techniques for satellite and satellite-terrestrial networks, respectively. 
%	The technique in ~\cite{lin2022slnr} uses signal-to-leakage-plus-noise ratio to optimize the beamforming vectors, which enhances the security and energy efficiency in satellite communications, but with ideal hardware.}
%Applying optimization techniques in~\cite{lin2022slnr,lin2021secrecy} to a hardware-impaired system is challenging due to highly-coupled IRS phase and beamforming matrices. The current study avoids this issue by designing a combiner that first maximizes the SINR, and later derives a closed-form solution for optimal phase and power.
 Recently, the authors in~\cite{bjornson2018hardware} developed a distortion aware minimum mean squared error (DA-MMSE) receiver for a single-cell mMIMO system without IRS.  Motivated by this idea, we propose a distortion-and-aging-aware MMSE (DAA-MMSE) receiver that mitigates the effect of hardware impairments and channel aging, in an IRS-assisted multi-cell mMIMO systems with Rician channels. The sum SE of a system can be further enhanced by optimally configuring the IRS phases~\cite{demir2022channel,nadeem2020asymptotic}.  Demir \textit{et al.} in~\cite{demir2022channel} designed optimal IRS phases based on instantaneous channel estimates. The authors in~\cite{nadeem2020asymptotic} designed IRS phase matrix using projected gradient ascent method. Both these works considered a \textit{single-cell IRS system, without channel aging, and with ideal hardware.} {We also note that existing multi-cell IRS works in \cite{pan2020multicell,buzzi2021ris} used \textit{instantaneous channels} to optimize the phase. {The current work, in contrast, proposes a novel IRS phase optimization framework that requires only channel statistics. 
% 		The optimal IRS phase, thus,  remain valid for a very long time interval.
 	}} We now list our \textbf{main contributions} work.\\
%\begin{itemize}
$\bullet$  We consider a hardware-impaired  multi-cell IRS mMIMO system with spatially-correlated Rician channels, and design a  DAA-MMSE receiver which mitigates the effects of hardware impairments and channel aging, and improves the SE.\\
$\bullet$  We develop a practically-implementable optimization algorithm to maximize the SE, which is a non-convex fractional function of  IRS phases. We achieve this objective by proposing a novel optimization framework that is based on the minorization-maximization (MM) approach~\cite{zhi2022ris}.\\ 
$\bullet$  We numerically show that the: i) SE gains of DAA-MMSE over DU-MMSE receiver increase with hardware impairment and channel aging; ii) DAA-MMSE receiver and IRS will reduce the increased pilot overhead due to channel aging;~iii) {IRS deployed at the cell edge of a multi-cell system~reduces the SE-per-user.} We also show the SE gains of  proposed phase optimization algorithm over its random phase counterpart. 
%\end{itemize}
%\vspace{-7pt}
\section{IRS assisted multi-cell mMIMO system model}%\vspace{-5pt}
We consider the uplink of an IRS-assisted multi-cell mMIMO system with $L$ cells. The BS is equipped with a uniform planar array (UPA), which has $N_H$ (resp. $N_V$) antennas per row (resp. column), with a total of $N=N_HN_V$ antennas. Each BS serves $K$ single-antenna UEs  in its respective cell. The UEs have a weak direct link with their BS due to high path-loss~\cite{demir2022channel}. To support such UEs, each cell also has IRS with $M = M_H M_V$ reflecting elements. %Due to the presence of the IRS, the signal received by the BS is a combination of signal received directly from UEs and reflected from the IRS.
 To reduce the hardware cost and power consumption, the BS and UEs are equipped with low-cost, hardware-impaired RF chains and low-resolution ADC/DACs. The communication takes place over a resource block of $\tau_c$ time instants, which is divided into uplink training period and data transmission intervals of length $\tau_p$ and $(\tau_c-\tau_p)$ time instants, respectively. Due to mobility of UEs, the channel varies over time in a correlated manner within the resource block~\cite{papazafeiropoulos2023impact}. We model these temporal channel variations by using the Jakes model~\cite{papazafeiropoulos2023impact}. We first model the UE-BS, UE-IRS and IRS-BS channels.
 \vspace{-0.1cm}
\subsection{{Channel model}}
%\subsection{Channel model} \label{dist}
%\vspace{-5pt}
%Due to the presence of IRS, the uplink signal received by the BS is a combination of the signal received directly from the UE, and the signal reflected by the IRS. 
Let $U_{lk}$ be the $k$th UE in $l$th cell. The effective channel between the UE $U_{lk}$ and the BS $j$ at the time instant $\lambda$, which is denoted as $\gv_{lk}^j[\lambda]$, consists of the direct UE-BS channel $\hv_{lk}^j[\lambda]$, and indirect UE-IRS-BS channels. The channel $\gv_{lk}^j[\lambda]$ is, accordingly, modeled as follows:
\begin{align}\label{comb_channel}
	\gv_{lk}^j[\lambda] = \hv_{lk}^j[\lambda]+\sum\limits_{i=1}^L\Xmat_i^j\Thetamat_i\zv_{lk}^i[\lambda].
\end{align}
The vector $\zv_{lk}^i[\lambda]$ denotes the channel between the UE $U_{lk}$ to the $i$th IRS. The matrix $\Xmat_i^j$ denotes the LoS channel between the IRS and BS, and is modelled similar to~\cite{nadeem2020asymptotic} as a high-rank channel. The matrix $\Thetamat_i$ is the phase-matrix of the IRS in $i$th cell, given as $\Thetamat_i = \text{diag}\big[e^{j\theta_{i,1}},\cdots,e^{j\theta_{i,M}}\big]$. The scalar $\theta_{i,m} \in [-\pi,\pi]$ is the phase of the $m$th element in the $i$th IRS.\\
\underline{\textit{Modeling UE-BS and UE-IRS channels}:} Due to insufficient antenna/element spacing at the BS/IRS and the presence of LoS links, the UE-BS channel $\hv_{lk}^j[\lambda]$ and UE-IRS channel $\zv_{lk}^i[\lambda]$ follow spatially-correlated Rician distribution~\cite{demir2022channel}. We model the channels $\{\hv_{lk}^j[\lambda],\zv_{lk}^i[\lambda]\}$ by referring them together $\tv_{lk}^q[\lambda] = \{\hv_{lk}^j[\lambda],\zv_{lk}^i[\lambda]\}$ with $q\in (j,i)$ as follows:
%. The channel, similar to ~\cite{demir2022channel}, is modeled as follows:
\begin{align}
	\tv_{lk}^j[\lambda] &=\overline{\tv}_{{lk}}^qe^{j\varphi_{t_{lk}}^q[\lambda]}+ \overline{\Rmat}_{t_{lk}}^{q\frac{1}{2}}\widetilde{\tv}_{lk}^q[\lambda].
\end{align}
Here $\overline{\tv}_{lk}^q = \sqrt{\frac{\beta_{t_{lk}}^q\kappa_{t_{lk}}^{q}}{1+\kappa_{t_{lk}}^{q}}}\breve{\tv}_{{lk}}^q$, $ \overline{\Rmat}_{t_{lk}}^q = {\frac{\beta_{t_{lk}}^q}{1+\kappa_{t_{lk}}^{q}}}\Rmat_{t_{lk}}^q$. The scalars $\kappa_{t_{lk}}^{q}$ and $\beta_{t_{lk}}^q$ represent the Rician factor and the large scale fading coefficient of the channel $\tv_{lk}^q$. The matrix $\Rmat_{t_{lk}}^q$ models the spatial correlation of the channel $\tv_{lk}^q$~\cite{demir2022channel}. The vector~$\breve{\tv}_{{lk}}^q$ denotes the LoS component. The scalar $\varphi_{t_{lk}}^q[\lambda]$ at the time instant $\lambda$ denotes the random phase shift in the LoS component, and is uniformly distributed between $[-\pi,\pi]$. 
%The existing studies in \cite{pan2020multicell,hua2020intelligent} have assumed a static LoS phase.  A slight change in UE position, however,  significantly modifies the phase~\cite{demir2022channel}. It is, therefore, crucial to consider this LoS phase shift~\cite{demir2022channel}.  
The vector $\widetilde{\tv}_{lk}^q[\lambda]$ denotes the small scale fading, which has probability density function (pdf) $\mathcal{CN}(\mathbf{0},\Imat_N)$. The  UE mobility causes the UE-BS and UE-IRS channels to vary across different time instants in a resource block, which leads to channel aging. To analyze it, we model the effective channel $\gv_{lk}^j[n] $ at the $n$th time instant, with $1 \leq n \leq \tau_c$, as a combination of channel at $\lambda$th time instant $\gv_{lk}^j[\lambda]$ and an innovation component $\qv_{lk}^j[n]$~as~\cite{papazafeiropoulos2023impact}:
\begin{align}
\gv_{lk}^j[n] = {\vartheta}_{lk}[\lambda-n]\gv_{lk}^j[\lambda]+\overline{\vartheta}_{lk}[\lambda-n]\qv_{lk}^j[n].
\end{align}
Here ${\vartheta}_{lk}[\lambda-n]$ is the temporal correlation, and $\overline{\vartheta}_{lk}[\lambda-n] = \sqrt{1-{\vartheta}_{lk}^2[\lambda-n]}$. Its value, based on Jakes model~\cite{papazafeiropoulos2023impact}, is given as~${\vartheta}_{lk}[\lambda-n] = J_0(2\pi f_{lk}^dT_sm)$.
For a UE velocity of $v_{lk}$, the Doppler shift is $f_{jk}^d = v_{lk}f_c/c$, with $f_c$ and $c$ being the carrier frequency and the velocity of light, respectively. The function $J_0(\cdot)$ is the zeroth-order Bessel function, and $T_s$ denote the sampling time~\cite{papazafeiropoulos2023impact,zhang2023channel}. The effective channel innovation component  $\qv_{lk}^j[n] = \overline{\hv}_{lk}^je^{j\varphi_{h_{lk}}^j[n]}+ \uv_{lk}^j[n]+\sum_{i=1}^L\Xmat_i^j\boldsymbol{\Theta}_i\big(\overline{\zv}_{lk}^ie^{j{\varphi}_{z_{lk}}^i[n]}+\vv_{lk}^i[n]\big) $ has zero mean and covariance $\Cmat_{g_{lk}^j} =\Cmat_{h_{lk}^j}+\sum_{i=1}^L\Xmat_i^j\boldsymbol{\Theta}_i\Cmat_{z_{lk}^i}\boldsymbol{\Theta}_i^H\Xmat_i^{j^H}$. Here, $\Cmat_{h_{lk}^j} = \overline{\hv}_{lk}^j\overline{\hv}_{lk}^j+\overline{\Rmat}_{h_{lk}}^j$, and $\Cmat_{z_{lk}^i} = \overline{\zv}_{lk}^i\overline{\zv}_{lk}^j+\overline{\Rmat}_{z_{lk}}^i$ are the covariance matrices of UE-BS and UE-IRS channels, respectively. The vectors $\uv_{lk}^j[n]$ and $\vv_{lk}^i[n]$ are the innovation components of the UE-BS and UE-IRS channels, and are distributed as $\mathcal{CN}(\mathbf{0},\overline{\Rmat}_{h_{lk}^j})$ and $\mathcal{CN}(\mathbf{0},\overline{\Rmat}_{z_{lk}^i})$, respectively. 
\begin{figure*}
%	\vspace{.13in}
	\begin{align}\label{rx_pilot_signal}
		\yv_p^j[t_k] &= \Amat^{j}\sum\limits_{l=1}^{L}\Big(\vartheta_{lk}[\lambda-t_k]\hv_{lk}^{j}[\lambda] + \overline{\vartheta}_{lk}[\lambda-t_k]\uv_{lk}^{j}[t_k]+ \sum\limits_{i=1}^{L}\Xmat_{i}^j\Thetamat_{i}\big({\vartheta}_{lk}[\lambda-t_k]\zv_{lk}^{i}[\lambda]+\overline{\vartheta}_{lk}[\lambda-t_k]\vv_{lk}^{i}[t_k]\big)\Big)\nonumber\\[-5pt]
		&\times\big(\alpha_{u_{lk}}\sqrt{\tilde{p}_{lk}}{\phimat}[t_k] +n_{\text{DAC}_{lk}}[t_k]\!+\!\eta_{u_{lk}}[t_k]\big)+\Amat^{j}\etamat_{\text{BS}}^{j}[t_k]+\Amat^{j}\nv^{j}[t_k]+\nv_{\text{ADC}}^{j}[t_k].
	\end{align}
	\hrule\vspace{-0.23in}
\end{figure*}
\vspace{-0.4cm}
\subsection{{Channel estimation}}\vspace{-0.1cm}
%Most existing IRS works~\cite{alwazani2020intelligent,deshpande2022spatially,buzzi2021ris} used separate estimations for direct channel and cascaded channels. This approach requires $\tau_c(M+1)$ time instants to estimate the channel, resulting in high pilot overhead. Contrary to works~\cite{alwazani2020intelligent,buzzi2021ris} and~\cite{deshpande2022spatially}, we estimate the overall channel consisting of the direct and cascaded channels~\cite{papazafeiropoulos2021intelligent}. 
In the uplink training phase, the UE $U_{lk}$ transmits the pilot $\vphantom{\sum\limits^K}\sqrt{\widetilde{p}_{lk}}\phi_k[t_k]$ at time instant $t_k$, with $|\phi_k[t_k]|^2 \!=\!1$. The term ${\widetilde{p}_{lk}}$ is pilot transmit power. We assume, similar to~\cite{buzzi2021ris}, that UEs with the same index in different cells transmit pilots at the same time index, which causes pilot contamination (PC). 
The UE $U_{lk}$ feeds its pilot signal to the low-resolution DAC, which distorts it. The distorted pilot, based on the Bussgang model~\cite{papazafeiropoulos2021intelligent}, is given as follows: 
\begin{align}
	s_{p_{lk}}[t_k] = \alpha_{u_{lk}}\sqrt{\tilde{p}_{lk}}{\phimat}[t_k] +n_{\text{DAC}_{lk}}[t_k].
\end{align}
Here $\alpha_{u_{lk}} = (1-\rho_{u_{lk}})$ is the Bussgang gain with $\rho_{u_{lk}}$ being the DAC distortion factor. The scalar $n_{\text{DAC}_{lk}}$ is the quantization noise with zero mean and variance $\rho_{u_{lk}}\alpha_{u_{lk}}\tilde{p}_{lk}$. The quantization noise is uncorrelated with the pilot signal $\sqrt{\tilde{p}_{lk}}\mathbf{\phi}(t_i)$. The output is fed to hardware-impaired RF chains, which add additive distortion noise $\eta_{u_{lk}}$ as follows:\vspace{+0.1pt}
\begin{align}
	\widetilde{s}_{p_{lk}}[t_k] = \alpha_{u_{lk}}\sqrt{\tilde{p}_{lk}}\phimat[t_k] +n_{\text{DAC}_{lk}}[t_k]+\eta_{u_{lk}}[t_k].
\end{align}
 The noise  $\eta_{u_{lk}}[t_k]$  according to the error vector magnitude (EVM) model~\cite{papazafeiropoulos2021intelligent}, has pdf $\mathcal{CN} (0,\kappa_{u}^2\delta_{u_{lk}})$, with $\delta_{u_{lk}} =\tilde{p}_{lk}\alpha_{u_{lk}}$. Here, $\kappa_u$ represents the UE transmit RF chain EVM, which is specified in the design data sheet~\cite{papazafeiropoulos2021intelligent}
The signal received at the BS antennas at the time instant $t_k$ is the sum of pilot signals transmitted from the UEs in all the cells i.e., 
\begin{align}
	\breve{\yv}^{j}_p[t_k] = \sum_{l=1}^{L}\gv_{lk}^{j}[t_k]\widetilde{s}_{p_{lk}}[t_k].
\end{align}
The received pilot signal at the $j$th BS antenna is fed to its hardware-impaired RF chain. The distorted RF output,  based on the EVM model~\cite{papazafeiropoulos2021intelligent}, is given as follows:
\begin{align}
	\widetilde{\yv}^j_p[t_k] = \sum_{l=1}^{L}\gv_{lk}^{j}[t_k]\widetilde{s}_{p_{lk}}[t_k]+\etamat_{\text{BS}}^{j}[t_k]+\nv^{j}[t_k].
\end{align}
The RF impairments $\etamat_{\text{BS}}^{j}[t_k]$ has pdf $\mathcal{CN} (\mathbf{0}_N,\kappa_{b}^2\Dmat^{j})$. The scalar $\kappa_{b}$ is the receive EVM, and matrix  $\Dmat^{j}\!\!\!=\!\!\!\text{diag}\{\E[\widetilde{\yv}_p^{j}[t_k]\widetilde{\yv}_p^{jH}[t_k]|\gv_{lk}^{j}]\}$.
%\!\!\! =\!\!\!  \sum\limits_{l= 1}^L\tilde{p}_{lk}\alpha_{u_{lk}}(1+\kappa_u^2)\text{diag}(\gv_{lk}^j[t_k]\gv_{lk}^{jH}[t_k])$.  
The vector $\nv^j[t_k]$ is the additive white Gaussian noise (AWGN), with $\mathcal{CN}(0,1)$ entries.  The $j$th BS feeds the distorted RF output to its low-resolution ADC, which introduces quantization errors. The distorted output, based on the Bussgang model~\cite{papazafeiropoulos2021intelligent}, is given as follows:
\begin{align}\label{pilot_signal}
	\yv_p^j[t_k] = \mathcal{Q}(\widetilde{\yv}^j_p[t_k]) = \Amat^{j}\widetilde{\yv}_p^j[t_k]+\nv_{\text{ADC}}^{j}[t_k].
\end{align}
The matrix $\Amat^{j} = \text{diag}\{1-\rho_{b_1}^{j}\cdots 1-\rho_{b_N}^{j}\}$, where $\rho_{b_N}^{j}$ models the ADC distortion. The vector $\nv_{\text{ADC}}^{j}[t_k]$, with pdf $\mathcal{CN}(\mathbf{0},\Tmat^{j}\Cmat^{j})$, represents the quantization noise, and is uncorrelated with $\widetilde{\yv}_p^j$ \cite{papazafeiropoulos2021intelligent}. Here $\Tmat^{j} = \Amat^{j}(\Imat_N -\Amat^{j})$ and $\Cmat^{j} = \text{diag}(\E[\widetilde{\yv}_p^{j}\widetilde{\yv}_p^{jH}]\big|\gv_{lk}^{j}[t_k])$. Recall that the channels between two different time instants are correlated. The received signal $\yv_p^j[t_k]$, thus, can be used to estimate channels at any instant $1 <n<\tau_c$ in the resource block. The estimate quality, however, deteriorates as the time difference between the pilot transmission $(1<n <\tau_p)$ and the estimation $(\tau_p+1<n<\tau_c)$ increases. We, therefore, estimate the channel at the time instant $\lambda= \tau_p +1$, and use these estimates to design the BS receiver. We now express the received pilot signal  $\yv_p^j[t_k]$  in terms of the channel at time instant $\lambda$ as in \eqref{rx_pilot_signal}, shown at the top of this page.
%\begin{align}\label{pilot_receive_signal}, where
By using~\eqref{rx_pilot_signal}, we estimate the effective channel $\gv_{lk}^j[\lambda]$ in the theorem below. It is proved in~\cite[Sec.~I]{proof_IRS}. Table~\ref{table:symbols} summarizes the notations used in~paper.
% As the BS is unaware of the phase-shifts, the estimate is a phase-unaware estimate.
%\vspace{-0.8cm}
\begin{theorem}\label{theo:channel_est}
	The LMMSE estimate of an IRS-assisted multi-cell mMIMO system with imperfect hardware and spatially-correlated Rician-faded channels with aging  is given as  
	\begin{align}\label{eq:ch_est}
		\widehat{\gv}_{lk}^j[\lambda] = \sqrt{\widetilde{p}_{lk}}\alpha_{u_{lk}}\vartheta_{lk}[\lambda-t_k]\Cmat_{\gv_{lk}^{j}}\Amat^{jH}\Psimat_{jk}^{-1}\yv_p^{j}[t_k]. 
	\end{align}
\end{theorem}\vspace{-5pt}
Here 
$\Psimat_{jk} = \sum_{l=1}^{L}\widetilde{p}_{lk}\alpha_{u_{lk}}^{2}\Amat^{j}\Cmat_{\gv_{lk}^{j}}\Amat^{jH}+ \sum_{l=1}^{L}\widetilde{p}_{lk}\alpha_{u_{lk}}\times $\\$(\rho_{u_{lk}}\!+\kappa_u^2)\Amat^{j}\Cmat_{g_{lk}^{j}}\Amat^{jH}
\!+\kappa_b^2\Amat^{j}\Dmat^{j}\Amat^{jH}\!+\sigma_b^2\Amat^{j}\Amat^{jH}+\Tmat^j\Cmat^j$.
% The covariance matrices $\Cmat_{g_{lk}^j}= (\overline{\hv}_{lk}^{j}(\overline{\hv}_{lk}^{j})^H+\Rmat_{h_{lk}^{j}})+ \sum\limits_{i=1}^{L} \Xmat_{i}\Thetamat_{i}\Cmat_{z_i}\Thetamat_{i}^{H}\Xmat_{i}^{H}$. 
The estimate $\widehat{\gv}_{lk}^j$ has the covariance matrix  $\Cmat_{\widehat{g}_{lk}^j} = \alpha_{u_{lk}}^2\widetilde{p}_{lk}\vartheta_{lk}^2[\lambda-t_k]\Cmat_{g_{lk}^{j}}\Amat^{jH}\Psimat_{jk}^{-1}\Amat^j\Cmat_{g_{lk}^{j}}$.
%==================
%\vspace{0.5in}
\begin{table}[t] \footnotesize
	\vspace{-.06cm}
	\centering
	\caption{{List of symbols}} \label{table:symbols}\vspace{-0.2cm}%\footnotesize
	{\begin{tabular}{ |c|c|}
			\hline
			\textbf{Symbol} & \textbf{Description} \\
			\hline
%			$L$, $M$, $N$ & Number of cells, IRS elements and BS antennas.\\
%			\hline
			$\alpha_{u_{lk}}$/$\alpha_{b_i}^j$ 	& Bussgang gain at UE/BS.\\
			\hline
			$\rho_{u_{lk}}$/$\rho_{b_N}^j$      & DAC/ADC distortion factor.\\
			\hline
			$\kappa_u$/$\kappa_b$,${\widetilde{p}_{lk}}$,$p_{lk}$	& UE/BS EVM, pilot power, UE transmit power.\\
			\hline
			$s_{jk}$/$\widetilde{s}_{lk}$	& Transmit data symbol/ RF chain output at UE.\\
			\hline
			$\breve{\yv}^{j}$/$\widetilde{\yv}^j$/${\yv}^j$	& {Received signal/RF chain output/ADC output at BS.}\\
			\hline
%			$u_{jk,n}$, $d_{jk,n}$	& Optimization-related auxiliary variables.\\
%			\hline
	\end{tabular}}
	 \vspace{-0.05cm}
\end{table}
%==================
% The estimation error $\widetilde{\gv}_{lk}^j = \gv_{lk}^j-\widehat{\gv}_{lk}^j$ has covariance matrix $\Cmat_{\widetilde{g}_{lk}^j} = \Cmat_{g_{lk}^j} - \Cmat_{\widehat{g}_{lk}^j}$. 
%Due to non-Gaussian nature of Rician-faded channel with random phase-shifts, the estimate and the estimation error are uncorrelated, but not independent.
%\newline
%\vspace{-0.5cm}
\subsection{{Data Transmission}}\vspace{-0.1cm}
At the $n$th instant of data transmission interval, the UE transmit its signal $\sqrt{p_{lk}}x_{lk}[n]$, with $\E|\sqrt{p_{lk}}x_{lk}|^2=p_{lk}$. The symbol is fed to the low-resolution DAC, and then to hardware-impaired RF chains. Its distorted output, based on the Bussgang and EVM model~\cite{papazafeiropoulos2021intelligent}, is  
\begin{align}\label{eq:ue_tx_signal}
	\widetilde{s}_{lk}[n] = \alpha_{u_{lk}}\sqrt{p_{lk}}x_{lk}[n] +n_{\text{DAC}_{lk}}[n]+\eta_{u_{lk}}[n].
\end{align}
The quantization noise   $n_{\text{DAC}_{lk}}$ has zero mean and variance $\alpha_{u_{lk}}(1-\alpha_{u_{lk}})p_{lk}$. The  noise $n_{\text{DAC}_{lk}}$ is uncorrelated with the input signal $\sqrt{p_{lk}}x_{lk}[n]$. The additive RF impairment noise $\eta_{u_{lk}}[n]$ has pdf $\mathcal{CN}(0,\kappa_{u}^2\delta_{u_{lk}})$, with $\delta_{u_{lk}} =\alpha_{u_{lk}}p_{lk}$. The  $j$th BS receives the following signal $\breve{\yv}^j = \sum_{l=1}^{L}\sum_{k = 1}^K\gv_{lk}^j[n]s_{\text{RF}_{lk}}$. This signal is fed to the RF chain, whose output, based on the EVM model~\cite{papazafeiropoulos2021intelligent},~is 
\begin{align}\label{eq:bs_rf_out_put}
	\widetilde{\yv}^j[n] = \sum\limits_{l=1}^L\sum\limits_{k = 1}^{K}\gv_{lk}^j[n]\widetilde{s}_{{lk}}[n]+\etamat_{\text{BS}}^j[n]+\nv^j[n].
\end{align}
The vector $\etamat_{\text{BS}}^j[n]$ is the RF distortion noise, whose pdf is $\mathcal{CN}(\mathbf{0},\kappa_{b}^2\Dmat^j)$, where $\Dmat^j =\text{diag}\{\E[\widetilde{\yv}^j\widetilde{\yv}^{jH}|\gv_{lk}^j]\}$. The vector $\nv^j[n]$ is AWGN, with pdf~$\mathcal{CN}(\mathbf{0}, \Imat_N)$. The BS then feeds the RF chain output to its low-resolution ADCs, whose  noisy output, based on the Bussgang model~\cite{papazafeiropoulos2021intelligent},~is 
\begin{align}\label{eq:bs_adc_out_put}
	\yv^j = \mathcal{Q}(\widetilde{\yv}^j[n]) = \Amat^j\widetilde{\yv}^j[n]+\nv_{\text{ADC}}^j[n].
\end{align}
The matrix $\Amat^j = \text{diag}\{\alpha_{b_1}^j\cdots \alpha_{b_N}^j\}$, with $\alpha_{b_N}^j$ being the Bussgang gain for the $i$th antenna of $j$th BS. The vector $\nv_{\text{ADC}}^j[n]$ denotes the quantization noise added at the $j$th BS  at the $n$th time instant. It has zero mean and covariance $\Amat^j(\Imat_N -\Amat^j)\Cmat^j$, with $\Cmat^j = \text{diag}(\E[\widetilde{\yv}^j\widetilde{\yv}^{jH}]|\gv_{lk}^j)$.
The received signal at the $j$th BS after substituting~\eqref{eq:ue_tx_signal} and~\eqref{eq:bs_rf_out_put} in~\eqref{eq:bs_adc_out_put}, is given as
\begin{align}
	\yv^j[n]&\!=\!\Amat^j\sum\limits_{l=1}^L\!\sum\limits_{k=1}^{K}\!\gv_{lk}^j[n] \big(\!\alpha_{u_{lk}}\sqrt{p_{lk}}x_{lk}\!+\!n_{\text{DAC}_{lk}}[n]\!+\!\eta_{u_{lk}}[n]\big)\notag\\&+\Amat^j\etamat_{\text{BS}}^j[n]\!+\!\Amat^j\nv^j[n]\!+\!\nv_{\text{ADC}}^j[n]. 
\end{align}
To decode the symbol $x_{jk}$, the $j$th BS combines the received signal using a receiver $\vv_{jk} \in \mathbb{C}^{N\times 1}$ designed using channel estimates. The resultant combined signal showing different interference terms, is given in~\eqref{bs_com_signal} at the top of next page. %\vspace{-5pt}
\begin{table*}[b]\footnotesize \vspace{-5pt}
	\centering 
	\caption{\vspace{0.06in} Desired signal and interference terms of SINR expression.}\label{sinr_terms}\vspace{-10pt}
	\begin{tabular}{ | c | c |  c |} 
		\hline
		$\overline{\text{DS}}_{jk,n}=\big|\alpha_{u_{jk}}\vartheta_{lk}[\lambda\!-\!n]\sqrt{p_{lk}}\E\big[\!\Gamma_{jk,jk}[\lambda]\big]\big|^2$& $\overline{\text{MUI}}_{jk,n}= \sum\limits_{l=1}^{L}\sum\limits_{i\neq k}^{K}\alpha_{u_{li}}^2{p_{li}}\E\big[|\Gamma_{jk,li}[n]|^2\big]$ & $\overline{\text{NS}}_{lk,n} \!\!=\!\! \E\left[|\vv_{lk}^{lH}\Amat\nv^l[n]|^2]\right]$\\
		\hline
		$\overline{\text{PC}}_{jk,n} = \sum\limits_{l\neq j}^{L}\alpha_{u_{jk}}^2p_{lk} \E\big[|\Gamma_{jk,lk}[n]|^2\big]$ & $\overline{\text{DAC}}_{jk,n} = \sum\limits_{l=1}^{L}\sum\limits_{i=1}^{K}\E\big[|\Gamma_{jk,li}[n] n_{\text{DAC}_{li}}[n]|^2\big]$ &$\overline{\text{TRF}}_{lk,n} = \sum\limits_{l=1}^{L}\sum\limits_{i=1}^{K}\!\E\big[|\Gamma_{jk,li}[n]\eta_{u_{li}}[n]|^2\big]$\\
		\hline
		$\overline{\text{RRF}}_{jk,n} = \E\big[|{\vv}_{jk}^{H}[\lambda]\Amat^{j}\boldsymbol{\eta}_{\text{BS}}^{j}[n]|^2 \big]$ & $\text{ADC}_{jk,n} = \E\left[|\vv_{lk}^{lH}\nv_{\text{ADC}}^l[n]|^2\right]$ & $\Gamma_{jk,li}[n] = \vv_{jk}^{l H}\Amat^{j} \gv_{li}^{j}[n]$\\
		\hline
		\multicolumn{3}{|c|}{$\overline{\text{CA}}_{jk,n} = \alpha_{u_{lk}}\bar{\vartheta}_{lk}[\lambda-n]\sqrt{p_{jk}}\E\big[|{\vv}_{jk}^{{H}}[\lambda]\Amat^{j}\qv_{jk}^{j}[n]|^2\big]$, $\overline{\text{BU}}_{jk,n} = \alpha_{u_{jk}}{p_{jk}}\vartheta_{lk}^2[\lambda-n]\E\big[|\Gamma_{jk,jk}[\lambda]|^2\big]-\overline{\text{DS}}_{jk,n}$ }\\[-0.5pt]
%		\hline
%		\multicolumn{2}{|c|}{$\Cmat_{h} = \overline{\hv}\overline{\hv}^H+\Rmat_h$ and $\Cmat_{z_i} = \overline{\zv}_i\overline{\zv}_i^H+\Rmat_{z_i}$, $\Cmat_{g} = \Cmat_{h} + \sum\limits_{i=1}^{L} \Xmat_{i}\Thetamat_{i}\Cmat_{z_i}\Thetamat_{i}^{H}\Xmat_{i}^{H}$}\\[-0.5pt]
		\hline
	\end{tabular}\vspace{-5pt}
\end{table*}%\vspace{-5pt}
\vspace{-0.7cm}
\subsection{{BS receiver design}}\vspace{-0.1cm}
The conventional MMSE  receiver, referred to as distortion-unaware (DU)-MMSE\cite{bjornson2018hardware}, is attractive due to its interference cancellation capability. However, it cannot mitigate the distortion caused due to  non-ideal hardware and channel aging. Motivated by~\cite{bjornson2018hardware}, we now propose a distortion and aging-aware (DAA-MMSE) receiver in Proposition~\ref{receiver}, which is  proved in Appendix~\ref{DAA_COMBINER}.%\vspace{-8pt}
\begin{Proposition}\label{receiver}
	For the considered system, the DAA-MMSE receiver that mitigates the detrimental effect of imperfect hardware and channel aging is given as 
	$\vv_{jk}^j = (\Dmat_{jk}^j)^{-1}\widetilde{\cv}_{jk}^j$ with $\widetilde{\cv}_{lk}^l = \alpha_{u_{jk}}\vartheta_{jk}[\lambda-n]\sqrt{p_{jk}}\Amat^{j}\gv_{jk}^{j}[\lambda]$.
	{The matrix $\Dmat_{jk}^j$ depends on channel estimate~$\widehat{\gv}_{jk}^j[\lambda]$, and is given in Appendix~\ref{DAA_COMBINER}.}
\end{Proposition}
\iftoggle{BIG_EQUATION}{
	\begin{figure*}[t!]
		\normalsize
		% Store the current equation number.
		\setcounter{mytempeqncnt}{\value{equation}}
		%\setcounter{equation}{11}
		%\hrule
		\vspace{.13in}
		\vspace{-8pt}
		\begin{align}
			&\yv_{jk}[n]=	\underbrace{\alpha_{u_{jk}}\sqrt{p_{jk}}\vv_{jk}^{H}\Amat^j\gv_{jk}^j[n]x_{jk}}_{\text{Desired signal $\widetilde{\text{DS}}_{jk,n}$}}+\underbrace{\vv_{jk}^{H}\Amat^j\sum\limits_{l=1}^L\sum\limits_{i\neq k}^{K}\alpha_{u_{li}}\sqrt{p_{li}}\gv_{li}^jx_{li}}_{\text{Multi-UE-interference, $\text{MUI}_{jk,n}$}}+\underbrace{\vv_{jk}^{H}\Amat^j\sum\limits_{l \neq j}^L\alpha_{u_{lk}}\sqrt{p_{lk}}\gv_{lk}^jx_{lk}}_{\text{Pilot contamination, $\text{PC}_{jk,n}$}}+\underbrace{\vv_{jk}^{H}\Amat\nv^j[n]}_{\text{AWGN noise at BS, $\text{NS}_{jk,n}$}}\notag\\[-4pt]&+
			\underbrace{\sum\limits_{l=1}^L\sum\limits_{i=1}^K\vv_{jk}^{H}\Amat^j\gv_{li}^j[n]n_{\text{DAC}_{li}}[n]}_{\text{DAC impairments at UE , $\text{DAC}_{jk,n}$}}+\underbrace{\sum\limits_{l=1}^L\sum\limits_{i=1}^K\vv_{jk}^{H}\Amat^j\gv_{li}^j[n]\eta_{u_{li}}[n]}_{\text{RF impairments at UE, $\text{TRF}_{jk,n}$}}\!+\!\!\!\underbrace{\vv_{jk}^{H}\Amat^j\etamat_{\text{BS}}^j[n]}_{\text{RF impairments at BS, $\text{RRF}_{jk,n}$}}\!\!\!+\!\!\!\underbrace{\vv_{jk}^{H}\nv_{\text{ADC}}^j[n]}_{\text{ADC impairments at BS, $\text{ADC}_{jk,n}$}}.\label{bs_com_signal}
		\end{align}
		\hrule
		\vspace{-0.28in}
\end{figure*}}{}\vspace{-5pt}
%\vspace{-0.2cm}
\section{Spectral efficiency analysis}%\vspace{-5pt}
We now exploit the use-and-then-forget (UatF) technique to derive a lower bound on the SE. 
%This technique is valid for mMIMO systems with sufficient channel hardening i.e., $\frac{1}{M}\widehat{\gv}_{jk}^{H}[\lambda]\Amat^j\gv_{jk}^j[\lambda] \approx \frac{1}{M}\E[\widehat{\gv}_{jk}^{H}[\lambda]\Amat^j\gv_{jk}^j[\lambda]]$. 
Using UatF technique, we decompose the desired signal $\widetilde{\text{DS}}_{jk,n}$ in~\eqref{bs_com_signal} as follows:
\begin{align}\label{uatf}
	y_{jk}[n]\!=	\!\alpha_{u_{jk}}\!\sqrt{p_{jk}}{\vartheta}_{lk}[\lambda\!-\!n]\E\big[\vv_{jk}^{H}\Amat^j\gv_{jk}^j[\lambda]\big]x_{jk} \!+\! w_{jk}.
\end{align}
The effective noise $w_{jk}$ contains all the terms in~\eqref{bs_com_signal} except the first term, and two extra terms, namely, beamforming uncertainty ${\text{BU}}_{jk,n} = \alpha_{u_{jk}}\sqrt{p_{jk}}{\vartheta}_{lk}[\lambda-n]|\vv_{jk}^{jH}\Amat^j\gv_{jk}^j[\lambda]-\E\big[\vv_{jk}^{jH}\Amat^j\gv_{jk}^j[\lambda]\big]$ and the channel aging term ${\text{CA}}_{jk,n} = \alpha_{u_{lk}}\bar{\vartheta}_{lk}[\lambda-n]\sqrt{p_{jk}}{\vv}_{jk}^{{H}}[\lambda]\Amat^{j}\qv_{jk}^{j}[n]$. 
%These are obtained after expressing the desired signal as ${\widetilde{\text{DS}}}_{jk,n} = {\text{DS}}_{jk,n}+{\text{BU}}_{jk,n}+{\text{CA}}_{jk,n}$. The modified signal term ${\text{DS}}_{jk,n}$ is used for the data detection is independent of instantaneous channels and depends on long-term channel statistics. 
The beamforming uncertainty ${\text{BU}}_{jk,n}$ denotes the signal received over an unknown channel. The ${\text{CA}}_{jk,n}$ term is obtained by expressing the combined channel $\gv_{jk}^j[n]$ at time instant $n$ as a combination of channel  $\gv_{jk}^j[\lambda]$ at the time instant $\lambda$, and its innovation component $\qv_{jk}^j[n]$. We note that the first term in \eqref{uatf} is uncorrelated with the effective noise term. Using \eqref{uatf}, the lower bound on the sum SE per cell of the system is~\cite{papazafeiropoulos2021intelligent}:
\begin{align}
	\text{SE}_{\text{sum}}= \frac{1}{L\tau_c}\sum\limits_{n= \lambda}^{\tau_c}\sum\limits_{j = 1}^L \sum\limits_{k = 1}^K\text{log}_2\left(1+\frac{\Delta_{jk,n}}{\Lambda_{jk,n}}\right), \text{ where }\label{sum_se_gen}
\end{align}
\begin{align*}
	\frac{\Delta_{jk,n}}{\Lambda_{jk,n}} \!=\! \frac{\overline{\text{DS}}_{jk,n}}{\begin{Bmatrix}\overline{\text{CA}}_{jk,n}\!+\!\overline{\text{BU}}_{jk,n}\!+\!\overline{\text{MUI}}_{jk,n}\!+\!\overline{\text{PC}}_{jk,n}\!+\!\overline{\text{DAC}}_{jk,n}\\ \text{TRF}_{jk,n}+\overline{\text{RRF}}_{jk,n}+\overline{\text{NS}}_{jk,n}+\overline{\text{DAC}}_{jk,n}\end{Bmatrix}}.
\end{align*}
The terms in the SE expression are given in Table~\ref{sinr_terms}, which are evaluated numerically through simulations. The lower-bound in~\eqref{sum_se_gen} is valid for any BS receiver.
\section{Low-complexity phase optimization algorithm}
%\vspace{-0.4cm}
We now design the optimal IRS phase matrix by maximizing the sum SE. The proposed solution depends only on the  channel statistics, which remain constant for $100$s of resource blocks, even for aging channels. 
%The IRS phases can be easily estimated at the BS before the beginning of actual communication. These phases are then used for estimating the composite channel. Since the IRS phase remains constant, the estimated channel also remains accurate for a long time. 
%We obtain the statistical-based phase optimization scheme by using the sum SE expression in~\eqref{sum_se_gen}.
 To maximize the SE at the $n$th transmission time instant, we design the IRS phase matrix $\Thetamat_{i}~\forall~i$ at time instant~$n$. The SE maximization problem, by ignoring the scalar $1/(\tau_c L)$ in \eqref{sum_se_gen} can be cast as follows:
\begin{subequations}
	\begin{alignat}{2}
		\mathbf{P1}:&
		\underset{\boldsymbol{\theta}}{\mbox{Max }} &&  \sum\limits_{n = \lambda}^{\tau_c}\!\sum\limits_{j=1}^L\!\sum\limits_{k = 1}^{K}\!\text{log}_2\Bigg(1\!+\!\frac{\Delta_{jk,n}(\boldsymbol{\theta})}{\Lambda_{jk,n}(\boldsymbol{\theta})}\Bigg)\label{se_obj},\\
		&\quad\text{s.t. }
		&& |[\boldsymbol{\theta}]_e| = 1~\forall~j,k,e.\label{P1_constraint}%\label{P2_constraint}
		%\label{problem0_1}
	\end{alignat}
\end{subequations}
 The vector $\boldsymbol{\theta} = [\boldsymbol{\theta}_{1},\cdots,\boldsymbol{\theta}_{L}]$ contains phases of all $L$ IRSs. Here $\boldsymbol{\theta}_{i} \delequal {vec}(\Thetamat_{i})$, where $vec(\cdot)$ denotes the vectorization operator. The constraint enforces unity modulus on each IRS element. 
We now solve $\mathbf{P1}$ which has following \textbf{optimization-related challenges:}
\begin{itemize}
	\item[C1.] The objective is a logarithmic function of scalar ratios in optimization variable $\boldsymbol{\theta}$. This makes $\mathbf{P1}$ a non-convex fractional programming problem.
	\item[C2.] The presence of IRS in each cell results in a sum SE, which is a coupled function of IRS phase matrices of all the cells. This inherent coupling introduces significant complexity in deriving the optimal solution $\boldsymbol{\theta}$.
\end{itemize}
 To handle challenge~C2, we restructure the SINR expression in terms of IRS phase $\boldsymbol{\theta}$ and deterministic matrices, which are given in~\cite{proof_IRS}. To address challenge C1, we develop an MM-based framework to handle the non-convex fraction, and then calculate the optimal $\boldsymbol{\theta}$. This approach provides a solution that requires only channel statistics.
 We begin by re-writing~$\mathbf{P1}$ as:
%\begin{subequations}
	\begin{align}
		\!\!\!\mathbf{P2}:
		\underset{\pv, \boldsymbol{\theta}}{\mbox{Max }}   \sum\limits_{n = \lambda}^{\tau_c}\sum\limits_{j=1}^L\sum\limits_{k = 1}^{K}f_{jk,n}(\boldsymbol{\theta}),\label{eq:phase_opt_obj}
		\:\:\text{s.t. }
		|[\boldsymbol{\theta}]_e| = 1~\forall e.
	%	\label{eq:problem_3b}
	\end{align}
%\end{subequations}
Here, $f_{jk,n}(\boldsymbol{\theta})\delequal \text{log}_2\Big(1+\frac{\boldsymbol{\theta}^{H}\Amat_{jk,n}\boldsymbol{\theta}}{\boldsymbol{\theta}^{H}\Bmat_{jk,n}\boldsymbol{\theta}}\Big)$. The matrices $\Amat_{jk,n}$ and $\Bmat_{jk,n}$ depend on the large-scale parameters, whose simplified expressions are given in~\cite{proof_IRS}. Problem~$\mathbf{P2}$ is non-convex as it contains sum of fractional ratios, with optimization variable $\boldsymbol{\theta}$ in its numerator and denominator. We solve it using the MM framework~\cite{zhi2022ris}, which considers the problem $\underset{\xv \in \mathcal{X} }{\mbox{Max }} a(\xv)$, The MM framework has  two steps.  In the first step, we find a surrogate function $f(\xv|\widehat{\xv}_t)$ that approximates the objective function $a(\xv)$. In the second step, we maximize the surrogate function $\widehat{\xv}_{t+1} = \underset{\xv}{\mbox{argmax }}f(\xv|\widehat{\xv}_t)$.
We now construct a novel surrogate for the objective $f_{jk,n}(\boldsymbol{\theta})$ in problem~$\mathbf{P2}$ in the following proposition, which is proved in~\cite{proof_IRS}.
% We now solve problem~\textbf{P2b} using MM  framework. The MM procedure consists of two steps. In the first
%majorization step, we find a surrogate function that locally
%approximates the objective function with their difference minimized
%at the current point. In other words, the surrogate
%upperbounds the objective function up to a constant. Then
%in the minimization step, we minimize the surrogate function To solve this problem using MM framework, 
\begin{Proposition}\label{prop:surro_fun}
	For a feasible point ${\boldsymbol{\theta}}^m$, a lower-bound of $f_{jk,n}(\boldsymbol{\theta})$, using Taylor's first order approximation, is given~by	
	\begin{align}
		f_{jk,n}(\boldsymbol{\theta}) \!&\geq \!\underbar f_{jk,n}(\boldsymbol{\theta}/\boldsymbol{\theta}^m) = {J}_{jk,n} \!\!+\!2~\text{Re}\Big\{\!(\rv_{jk,n}^m)^H\boldsymbol{\theta}\!\Big\},~\!\text{where}\notag\\[-2pt]
		{J}_{jk,n} &\!=\! f_{jk,n}(\boldsymbol{\theta}^m)- \frac{\boldsymbol{\theta}^{mH}\Amat_{jk,n}\boldsymbol{\theta}^m}{\boldsymbol{\theta}^{mH}\Bmat_{jk,n}\boldsymbol{\theta}^m}
		-\overline{\beta}_{jk,n}\boldsymbol{\theta}^{mH}(\lambda^{\text{max}}_{jk,n}\Imat_N\notag\\&-(\Bmat_{jk,n}+\Amat_{jk,n}))\boldsymbol{\theta}^m-N\overline{\beta}_{jk,n}\lambda^{\text{max}}_{jk,n}, \notag \\[-2pt]
		(\rv_{jk,n}^m)^H &=\omega_{jk,n}\boldsymbol{\theta}^{mH}\Amat_{jk,n}-\overline{\beta}_{jk,n}\boldsymbol{\theta}^{mH}((\Bmat_{jk,n}+\Amat_{jk,n})\notag\\[-2pt]&-\lambda^{\text{max}}_{jk,n}\Imat_N),~\omega_{jk,n} = {1}/{(\boldsymbol{\theta}^{mH}\Bmat_{jk,n}\boldsymbol{\theta}^m)}, \label{eq:surrog}
	\end{align}
%	\begin{align}
		$\overline{\beta}_{jk,n} = {\boldsymbol{\theta}^{mH}\Amat_{jk,n}\boldsymbol{\theta}^m\omega_{jk,n}\overline{\omega}_{jk,n}}$,~ $\overline{\omega}_{jk,n} \!=\!1/ ( \boldsymbol{\theta}^{mH}(\Bmat_{jk,n}+\Amat_{jk,n})\boldsymbol{\theta}^m)$  and   $\lambda^{\text{max}}_{jk,n} = \text{max}\{\text{eig}(\Bmat_{jk,n}+\Amat_{jk,n})\}$
	\end{Proposition}
%	\end{align} 
The function in~\eqref{eq:surrog} can be shown to satisfy conditions~\cite[Eqs. (55)-(57)]{zhi2022ris}, and is therefore a valid surrogate function.
Problem~$\mathbf{P2}$ is now reformulated using the proposed surrogate function $\underbar f_{jk,n}(\boldsymbol{\theta}/\boldsymbol{\theta}^m)$ in Proposition~\ref{prop:surro_fun} as follows:
%\begin{subequations}
	\begin{align}
		\mathbf{P3}:
		\underset{\boldsymbol{\theta}}{\mbox{Max }}  \sum\limits_{n = \lambda}^{\tau_c}\text{Re}\big\{\bar{\rv}_n^{mH}\boldsymbol{\theta}\big\},
		\quad\text{s.t. } \eqref{P1_constraint}.
		%\label{problem0_1}
	\end{align}
%\end{subequations}
{The vector $\bar{\rv}_n^m = \sum_{j=1}^L\sum_{k = 1}^{K}\rv_{jk,n}^{m}$.} The scalar $J_{jk,n}$ in $\underbar f_{jk,n}(\boldsymbol{\theta}/\boldsymbol{\theta}^m)$ is independent of variable $\boldsymbol{\theta}$, and  is omitted in the objective of $\mathbf{P3}$. The vector $\boldsymbol{\theta}^m$ denotes the IRS phase in the $m$th iteration. For a given initial $\boldsymbol{\theta}^m$, we calculate $\rv_{jk,n}^m$ by using the Proposition~\ref{prop:surro_fun}. The value of $\boldsymbol{\theta}$ that maximizes the objective of $\mathbf{P3}$ must be in-phase with $\bar{\rv}_n^m$. {Accordingly, the optimal phase for $(m+\!1)$th iteration is given~as follows:}
\begin{align}\label{eq:phase_opt}
	{\boldsymbol{\theta}}^{(m+1)} 
	=\text{exp}\left\{j \angle\bar{\rv}_n^m\right\}.
\end{align}
\vspace{-10pt}
\begin{algorithm}[htbp]
	\footnotesize
	%\SetAlgoLined
	\KwInput{Given a tolerance $\epsilon_r$ and number of iterations $M$.}
	\KwOutput{$\boldsymbol{\theta}$ }
	\For{$i \gets 1$ \textbf{to} $M$}{
				Given $\pv$ and $\boldsymbol{\theta}^{m}$, find $\boldsymbol{\theta}^{(m+1)}$ using~\eqref{eq:phase_opt}.\\
				%			Initialize $\boldsymbol{\theta}_n^m = \boldsymbol{\theta}^{(m+1)'}_{n}$, and find $\boldsymbol{\theta}^{(m+1)''}_{n} = \boldsymbol{\theta}^{(m+1)}_{n}$ using~\eqref{eq:phase_opt}.\\
				%			Find $\Delta_{\boldsymbol{\theta}^n_1} = \boldsymbol{\theta}^{(m+1)'}_{n} - \boldsymbol{\theta}^{m}_{n}$,  $\Delta_{\boldsymbol{\theta}^n_2} = \boldsymbol{\theta}^{(m+1)''}_{n}-\boldsymbol{\theta}^{(m+1)'}_{n} - \Delta_{\boldsymbol{\theta}^n_1}$,
				%			$\rho = -\frac{\|\Delta_{\boldsymbol{\theta}^n_1}\|}{\|\Delta_{\boldsymbol{\theta}^n_2}\|}$, and obtain optimal $\boldsymbol{\theta}^{m+1}_{n} = -\text{exp}\{j\angle(\boldsymbol{\theta}_{m}^{n}-2\rho\Delta_{\boldsymbol{\theta}^n_1} + \rho^2\Delta_{\boldsymbol{\theta}^n_2})\}$.\\
				%			\While{$\boldsymbol{\theta}^{m+1}_{n}$ \textnormal{does not increase the value of objective of}~\eqref{phase_obj}}{
					%				$\rho = (\rho-1)/2$ and $\boldsymbol{\theta}^{m+1}_{n} = -\text{exp}\{j\angle(\boldsymbol{\theta}^{m}_{n}-2\rho\Delta_{\boldsymbol{\theta}^n_1} + \rho^2\Delta_{\boldsymbol{\theta}^n_2})\}$.\\
					%			}
				Do until convergence: \If{$||\boldsymbol{\theta}^{m+1}-\boldsymbol{\theta}^{m}||  <\epsilon_r$}{break.}
			}
			\caption{IRS phase optimization.}
			\label{algo:opt_algo}
			%	\vspace{-.3cm}
		\end{algorithm}
	%	\vspace{-10pt}
%		\begin{remark}
%			Note that the SE lower bound in~\eqref{sum_se_gen} is valid for any BS combiner. The proposed IRS phase optimization is thus applicable to MRC, DU-MMSE, and the proposed DAA-MMSE designs. 
%		\end{remark}
%\vspace{0.05in}

{\subsubsection*{Complexity of Algorithm~\ref{algo:opt_algo}} The computation of  $\boldsymbol{\theta}^{m+1}$ in Step-$2$ using~\eqref{eq:phase_opt}, depends on $\rv_{jk,n}^m$. 
	 The calculation of vector $\rv_{jk,n}^m$, as seen  from Proposition~\ref{prop:surro_fun}, involves matrix multiplications with a complexity of $\mathcal{O}(M^2L^2K)$, while the existing work in~\cite{pan2020multicell} has a complexity of $\mathcal{O}(M^2(M+1)L^2K)$}.
%-----------------------------------------------------------------
\vspace{-0.5cm}
\section{Simulation Results} \label{sec:Simulation_results}
\vspace{-0.1cm}
We now numerically evaluate the SE of the proposed DAA-MMSE receiver. For this study, we consider a four-cell mMIMO network in a square area of $0.5$ km $\times$ $0.5$ km, wrapped around its edges. Each BS is located at the cell-center, while the IRS is placed at a distance of $0.125$~km from BS. The UEs are randomly distributed within a $90^{\circ}$ sector. Due to severe path loss, the BS-UE direct channel is attenuated by $70$~dB~\cite{demir2022channel}. {The large-scale fading coefficients, correlation matrices and Rician factors for the UE-BS and UE-IRS channels are modelled as in~\cite{demir2022channel}}. We assume a UPA at the BS and IRS with $N = 64$ BS antennas and $M = 100$ IRS elements respectively, UE transmit power $p_{lk} = 20$~dBm and $K = 5$ UEs per cell moving with a velocity of $v = 72$ km/h. The RF impairments are set as $\kappa_b = 0.1$, $\kappa_u = 0.05$, and $b=4$ bit ADC/DAC resolution. \\%For ideal hardware RF impairments, $\kappa_b$ = $\kappa_u =0$,  and ADC/DAC bit resolution $b=\infty$.\\% These values remain fixed unless explicitly specified.
\begin{figure*}[htbp]\vspace{0.07in}
	\begin{subfigure}{.24\textwidth}
	\includegraphics[width=\linewidth]{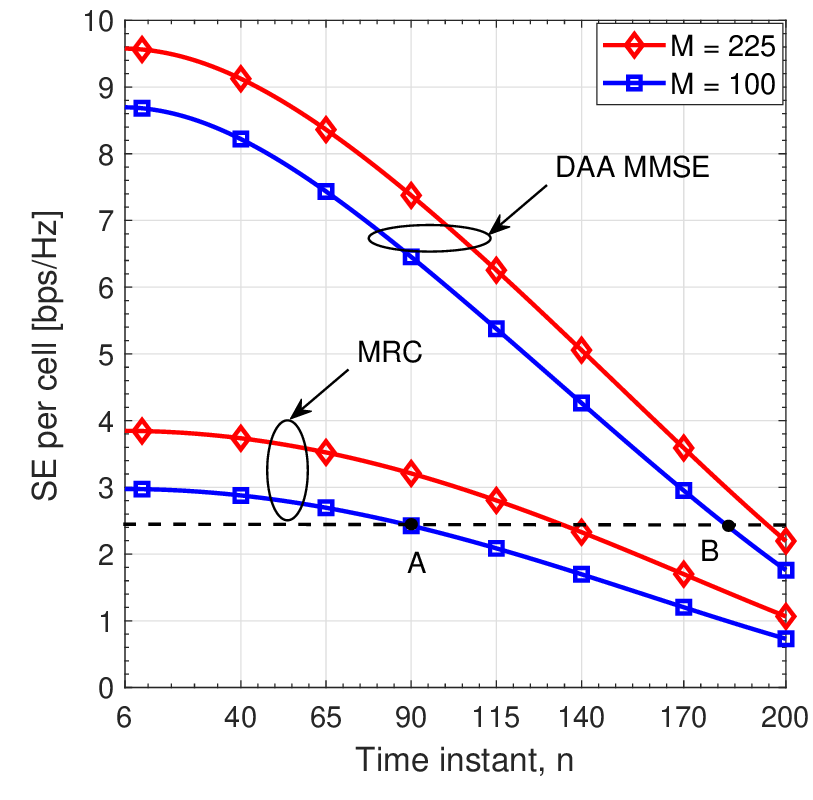}\vspace{-0.1cm}
	\caption{\small}
	\label{se_dtt}
\end{subfigure}\vspace{-10pt}
	\begin{subfigure}{.24\textwidth}
		\includegraphics[width=\linewidth]{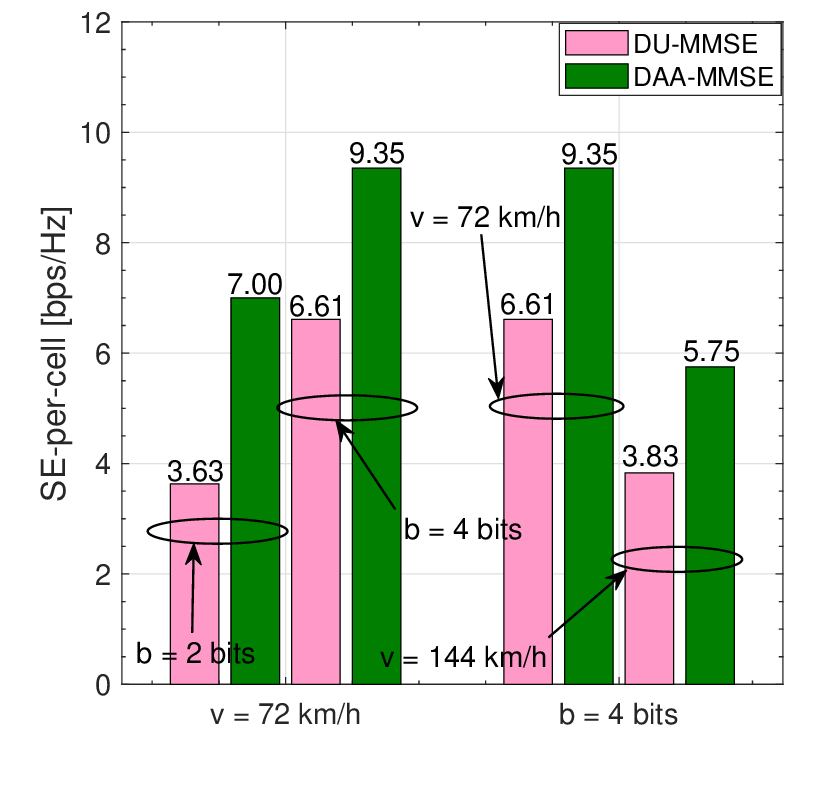}\vspace{-0.1cm}
		\caption{\small}
		\label{comb_comp}
	\end{subfigure}
	\begin{subfigure}{.24\textwidth}
	\includegraphics[width=\linewidth]{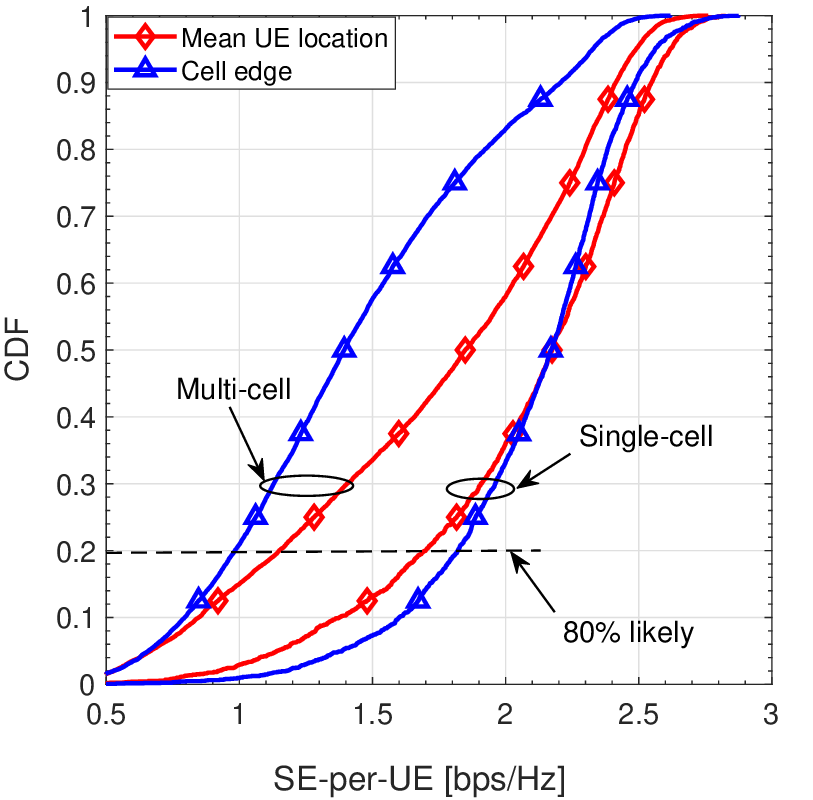}\vspace{-0.15cm}
	\caption{\small}
	\label{CDF_PLOT}
    \end{subfigure}
	\begin{subfigure}{.24\textwidth}
	\includegraphics[width=\linewidth]{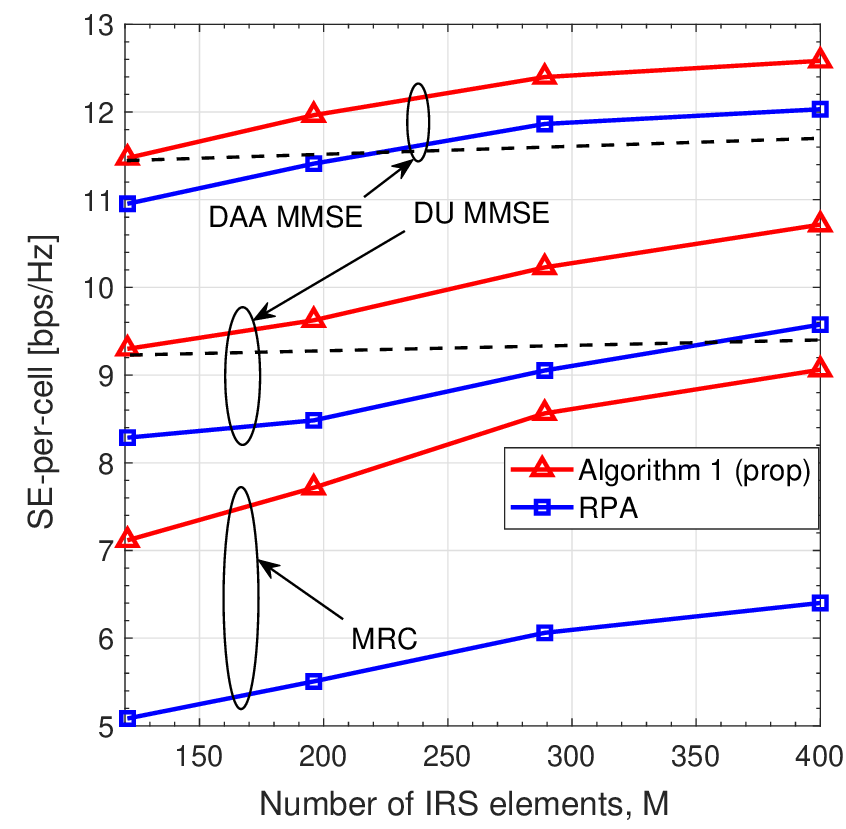}\vspace{-0.15cm}
	\caption{\small}
	\label{opt_1}
\end{subfigure}
	\caption{ a) DAA-MMSE vs MRC comparison; and  b) DAA-MMSE vs DU-MMSE receivers; c) CDF of  SE-per-UE for multi-cell and single-cell systems d) Comparison of proposed Algorithm with RPA for MR, DU-MMSE, and DAA-MMSE receivers. \\[-21pt]}
\end{figure*}
\hspace{-5pt}\underline{\textbf{Effect of channel aging}} We first compare in Fig.~\ref{se_dtt}, the instantaneous SE at each transmission time instant $n$ for the MRC and DAA-MMSE receivers. This study will help in determining the  transmission time instant for which a given QoS, specified in terms of SE, can be satisfied. We first observe that the SE reduces with increase in transmission time instant $n$. This is because the channel ages with $n$. We also see that the DAA-MMSE receiver outperforms MRC for all the time instants, which shows its effectiveness in handling the degradation due to channel aging. For $M=100$ IRS elements and a QoS requirement of $2.5$~bps/Hz, the DAA-MMSE receiver provides an SE $ \geq 2.5$~bps/Hz till $ n = 180$ (marked as point~B). The  MRC receiver provides it is only until $n = 90$ (marked as point~A). To maintain a QoS of $2.5$ bps/Hz, the BS should, therefore, re-estimate the channel after $n = 180$ (resp. $n = 90$) for the DAA-MMSE (resp. MRC) receiver. \textit{Our findings, thus, reveal that the proposed DAA-MMSE receiver reduces the pilot overhead required to obtain a desired QoS.} The QoS can also be maintained by increasing the IRS elements.  For example,  for MRC, QoS of $2.5$~bps/Hz can be achieved till $n =140$th time instant, but with $M =225$ IRS elements. We finally see  that for MRC, the SE degrades at a lower rate than the DAA-MMSE receiver. \\
\underline{\textbf{Comparison of DAA-MMSE and DU-MMSE receivers:}} We plot in Fig.~\ref{comb_comp} the SE obtained by both these receivers for a fixed: i) UE velocity of $v = 72$~km/h and two different ADC/DAC resolutions of $b = \{2,4\}$~bits; ii) $b =4$ bits and different UE velocities $v = \{72,144\}$~km/h. We observe that the DAA-MMSE receiver has a higher SE than DU-MMSE for all the cases. We also observe that the SE gain of DAA-MMSE receiver increases with hardware impairments and channel aging. For example, for UE speed of $v = 72$~km/h, for $b =4$ and $b =2$ is $41.45\%$, and $92.84\%$, respectively. For  $b=4$ bits, the SE gain  with UE velocity $v = 72$~km/h and $v = 144$~km/h is $41.45\%$ and $50.13\%$, respectively.\\
\underline{\textbf{Impact of IRS location on SE}} To study this aspect,  we now plot in Fig.~\ref{CDF_PLOT}, the cumulative distribution function (CDF) of SE-per-UE  for two cases i) Case-1: IRS is placed at cell edge, and ii) Case-2: IRS is placed at mean UE location. For this study, we consider DAA-MMSE receiver, and fix $M = 225$ IRS elements. We observe that the $80\%$ likely SE-per-UE in a single-cell system is higher for Case-1 because  IRS increases the signal strength of cell-edge UEs. However, the IRS at cell-edge in a multi-cell scenario leads to a lower $80\%$ likely SE-per-UE than the  IRS placed at mean UE locations. This is because the IRS, when placed at cell-edge, also boosts the MUI from other cells. \textit{This study informs a designer about  different IRS  placements for single- and multi-cell scenario.}\\
\underline{\textbf{SE optimization:}} We now investigate in Fig.~\ref{opt_1} the effectiveness of our SE optimization Algorithm \ref{algo:opt_algo}, when used with the DAA-MMSE, DU-MMSE, and MRC receivers. We compare its performance with random phase allocation (RPA), which randomly allocates phase in $[-\pi,\pi]$. We vary the number of IRS elements $M$ for this study.  We see that Algorithm~\ref{algo:opt_algo} outperforms RPA for all three receivers. Also, Algorithm~\ref{algo:opt_algo} marginally increases the  SE of DAA-MMSE receiver, when compared with RPA. \textit{This is because the DAA-MMSE receiver cancels the distortions due to hardware impairments and channel aging by using their statistical knowledge.} The Algorithm~\ref{algo:opt_algo}, therefore, only slightly helps in further suppressing them. The MRC and DU-MMSE receivers do not cancel these distortions. The Algorithm~\ref{algo:opt_algo} helps them in mitigating their effect by suitably adjusting the phase. 
\vspace{-6pt}
\section{Conclusion}\vspace{-4pt}
We proposed a low-complexity IRS phase optimization to maximize the SE, which requires only channel statistics. We  showed that our DAA-MMSE receiver outperforms MRC and DU-MMSE receivers, and reduces the channel estimation overhead required for aging channels. We also showed that the SE gain of the DAA-MMSE receiver over DU-MMSE receiver increases with hardware impairments and channel aging.  
%	We finally showed that IRS  also reduces the channel estimation overhead due to channel aging.
\appendices\vspace{-8pt}

\iftoggle{BIG_EQUATION}{
	\begin{figure*}[t!] \vspace{0.04in}
		\normalsize
		% Store the current equation number.
		\setcounter{mytempeqncnt}{\value{equation}}
		%\setcounter{equation}{11}
		%\hrule
%\begin{align*}
{\small	$\Dmat_{jk} = p_{jk}\alpha_{u_{jk}}^2\vartheta_{jk}^2[\lambda-n]\Amat^{l}\Cmat_{\widetilde{g}_{jk}^{j}}\Amat^{j^H}
	+p_{jk}\alpha_{u_{jk}}^2\bar{\vartheta}_{jk}^2[\lambda-n]\Amat^{j}\Cmat_{g_{jk}^{j}}\Amat^{j^H}+\sum\limits_{l=1}^{L}\sum\limits_{i\neq k}^{K}\alpha_{u_{ji}}^2p_{ji}\Amat^{l}(\vartheta_{li}^2[\lambda-n]\widehat{g}_{li}^j[\lambda]\widehat{g}_{li}^{jH}[\lambda]+\vartheta_{li}^2[\lambda-n]\Cmat_{\widetilde{g}_{li}^j}+ $\\[-3pt]$\bar{\vartheta}_{li}^2[\lambda-n]\Cmat_{g_{li}^j})\Amat^{j^H}
	+\sum\limits_{l\neq j}^{L}\alpha_{u_{lk}})^2p_{lk}\:\Amat^{j}(\vartheta_{jk}^2[\lambda-n]\widehat{\gv}_{li}^l[\lambda]\widehat{\gv}_{li}^{l^H}[\lambda]+\vartheta_{lk}^2[\lambda-n]\Cmat_{\widetilde{g}_{lk}^j}+\bar{\vartheta}_{lk}^2[\lambda-n]\Cmat_{g_{lk}^j})
	+\sum\limits_{l=1}^{L}\sum\limits_{i=1}^{K}\alpha_{u_{li}}(1\!-\!\alpha_{u_{li}})\widetilde{p}_{ji}\Amat^{j}(\vartheta_{lk}^2[\lambda\!-\!n]\widehat{\gv}_{li}^l[\lambda]\widehat{\gv}_{li}^{l^H}[\lambda]+\vartheta_{lk}^2[\lambda-n]\Cmat_{\widetilde{g}_{lk}^j}+\bar{\vartheta}_{lk}^2[\lambda-n]\Cmat_{g_{lk}^j})\Amat^{j^H}
	\!\!\!+\!\!\!\sum\limits_{l=1}^{L}\sum\limits_{i=1}^{K}\kappa_u^2\widetilde{p}_{ji}\alpha_{u_{ji}}\Amat^{l}(\vartheta_{lk}^2[\lambda\!-\!n]$\vspace{-2pt}
	\begin{align}\label{receiver_inv}
		\widehat{\gv}_{li}^j[\lambda]\widehat{\gv}_{li}^{j^H}[\lambda]+\vartheta_{lk}^2[\lambda-n]\Cmat_{\widetilde{g}_{lk}^j}+\bar{\vartheta}_{lk}^2[\lambda-n]\Cmat_{g_{lk}^j})
	+\kappa_b^2\Amat^{j}\Dmat^j\Amat^{j^H}+\Amat^{j}\Amat^j(\Imat_N-\Amat^j)\Cmat^j\Amat^{j^H}+\Amat^{j}\Amat^{j^H}, \Cmat_{\widetilde{g}_{lk}^j} = \Cmat_{{g}_{lk}^j}-\Cmat_{{g}_{lk}^j}.\end{align}}
%\end{align*}
		%\setcounter{equation}{10}
		\hrule
		\vspace{-0.2in}
\end{figure*}}{}
\section{}\label{DAA_COMBINER}
We derive the optimal DAA-MMSE receiver by considering the interference plus noise signal in~\eqref{bs_com_signal}, which is given as
{\small \begin{align*}
	&d_{jk} = \vv_{jk}^{H}\Amat^j\Bigg[\sum\limits_{l=1}^L\sum\limits_{i\neq k}^{K}\alpha_{u_{ji}}\sqrt{p_{ji}}\gv_{li}^j[n]x_{ji}\!+\!\etamat_{\text{BS}}^j[n]+\nv^j[n]\\[-2pt] &+
	\sum\limits_{l=1}^L\sum\limits_{i=1}^K(n_{\text{DAC}_{li}}[n]+\eta_{u_{li}}[n])+\sum\limits_{l\neq j}^L\alpha_{u_{jk}}\sqrt{p_{lk}}\gv_{lk}^jx_{lk}\Bigg]+\vv_{jk}^{H}\nv.
\end{align*}}\vspace{-1pt}
The BS has channel estimate $\widehat{\gv}_{jk}^j[\lambda]$, via which it can estimate channel of other time instances. It can, thus, find the conditional covariance $\rho_{d_{jk}} = \E[|d_{jk}|^2{\widehat{\gv}_{jk}^j[n]}]$ and the corresponding SINR at the $n$th time instant in terms of $\vv_{jk}$ as: $\text{SINR} = \frac{|\vv_{jk}^H\widetilde{\cv}_{jk}|^2}{\vv_{jk}^H\Dmat_{jk}^{-1}\vv_{jk}}$, with $\widetilde{\cv}_{jk} = \alpha_{u_{jk}}\vartheta_{jk}[\lambda-n]\sqrt{p_{jk}}\Amat^{j}\gv_{jk}^{j}[\lambda]$  and $\Dmat_{jk}$ shown  in~\eqref{receiver_inv}. By using Rayleigh coefficient theorem~\cite{bjornson2018hardware}, the optimal SINR-maximizing receiver is $\vv_{jk} =~\Dmat_{jk}^{-1}\widetilde{c}_{jk}$.
\bstctlcite{IEEEexample:BSTcontrol}
\vspace{-8pt}
\bibliography{IEEEabrv,DF_Ref}
\bibliographystyle{IEEEtran}
\end{document}